\documentclass[showpacs,twocolumn,amsmath,superscriptaddress,pre]{revtex4}
\usepackage{graphicx}
\usepackage{amsmath}

\usepackage{epsfig}
\usepackage{graphics}
\usepackage{latexsym}
\usepackage{amsfonts}
\usepackage{amssymb}
   
\usepackage{plain}    
  
\begin{document}

\title{Influence of the driving mechanism on the response of systems with athermal dynamics: the example of the random-field Ising model}

\author{Xavier Illa}
\affiliation{ Departament d'Estructura i Constituents de la Mat\`eria,
  Universitat de Barcelona \\ Diagonal 647, Facultat de F\'{\i}sica,
  08028 Barcelona, Catalonia}

\author{Martin-Luc Rosinberg}
%
%
\affiliation{Laboratoire de Physique Th\'eorique de la Mati\`ere Condens\'ee,
Universit\'e Pierre et Marie Curie, 4 Place Jussieu, 75252 Paris, France}

\author{Eduard Vives}
%
%
\affiliation{ Departament d'Estructura i Constituents de la Mat\`eria,
  Universitat de Barcelona \\ Diagonal 647, Facultat de F\'{\i}sica,
  08028 Barcelona, Catalonia}
%


\begin{abstract}
We investigate the influence of the driving mechanism on the hysteretic response of systems with athermal dynamics. In the framework of local-mean field theory at finite temperature (but neglecting thermallly activated processes), we compare the  rate-independent hysteresis loops obtained in the random field Ising model (RFIM) when controlling either the external magnetic field $H$ or the extensive magnetization $M$. Two distinct behaviors are observed, depending on disorder strength. At large disorder, the $H$-driven and $M$-driven protocols yield identical hysteresis loops in the thermodynamic limit. At low disorder, when the $H$-driven magnetization curve is discontinuous (due to the presence of a macroscopic avalanche), the $M$-driven loop is re-entrant while the induced field exhibits strong intermittent fluctuations and is only weakly self-averaging. The relevance of these results to the experimental observations in ferromagnetic materials, shape memory alloys, and other  disordered systems is discussed. 
\end{abstract}
\maketitle

\section{Introduction} 

When slowly driven by an external force, systems with quenched-in disorder often respond via a sequence of random bursts or avalanches  whose size distribution extends over several decades, ranging from microscopic to macroscopic. 
A classic example is the Barkhausen noise that occurs in soft magnetic materials as one varies the external magnetic field $H$. The random sequence of spikes observed in the induced voltage (that is, essentially, in the time-derivative of the induced magnetization $M$) is then related to the fluctuations in the motion of the magnetic domain walls due to structural disorder\cite{DZ2004}.   Avalanche dynamics is also observed in other systems, as for instance  in the acoustic emission associated to martensitic transformations\cite{VOMRPP1994} or during the drainage of superfluid $^4$He from a mesoporous solid\cite{LFH1993}.
All these systems also exhibit rate-independent hysteresis when the external force is cycled up and down, thermal fluctuations being negligible  on the experimental time scale due to the presence of very high energy barriers (the systems then explore a set of long-lived metastable states).

Avalanches are characteristic of a system driven by an
external force. In some experimental situations, however, it may be more convenient to control the conjugated {\it extensive} variable (or its time-derivative) rather than the external field. In magnetic materials for instance, some suitable feedback control may be used so to impose the flux (or its time derivative) instead of the magnetic field. This allows to study materials that reverse their magnetization very rapidly\cite{B1998}. In gas adsorption experiments, there are also cases where one controls the amount of gas introduced inside the porous material and measures the induced change in the pressure\cite{WC1990}. In shape memory alloys, stress-strain curves are routinely obtained by controlling the deformation of the sample and measuring the stress\cite{OW1998}. The following question then naturally arises: what is the relation between the force-response diagrams (magnetization curves, adsorption isotherms, strain-stress curves,etc...) observed in the two situations, say ``field-driven" and ``magnetization-driven" (hereafter, we shall adopt the language of magnetic systems).  This question is especially intriguing when there is  a macroscopic instability, e.g. a macroscopic jump in the magnetization curve obtained with the field-driven protocol. In this case, one may force a system that would spontaneously jump to a stable (or metastable) situation to remain into some other metastable or maybe unstable state. This may lead to a re-entrant behavior in the hysteresis loop, as indeed observed experimentally\cite{B1998,LS2002,BMPRV2006}. To what extent this behavior reveals an intrinsic property of the free-energy landscape of the system or depends on the experimental protocol is  not clear. To the best of our knowledge, this problem has not been seriously investigated in the literature, in spite of the intense theoretical activity focusing on out-of-equilibrium driven systems. 

In order to address this issue, we consider here a microscopic model, the Random Field Ising Model (RFIM), which has been extensively studied at zero temperature as a prototype of disordered hysteretic system with avalanche behavior\cite{SDP2004}. In particular, in dimension $3$, the  RFIM  is known to exhibit an out-of-equilibrium phase transition between a strong-disorder regime where the magnetization hysteresis loop is smooth on the macroscopic scale and a low-disorder one where it displays a macroscopic jump. The nonequilibrium  RFIM at $T=0$ is probably the simplest model to study the connection between metastability and hysteresis, and we expect its behavior in the magnetization-driven case to be relevant to a wide class of systems. However, as is discussed elsewhere\cite{IRSV2006}, the definition of a metastable $M$-driven dynamics for Ising spins at $T=0$ is somewhat problematic, and it is more convenient to work with continuous variables. This can be done in various ways and here we choose to study the RFIM at {\it finite} temperatures, using a local mean-field approach in which each Ising spin is replaced by its thermal average (see e.g. Ref.\cite{SLG1983}). One then  follows the evolution of the  system among the local minima of the free-energy landscape, discarding all thermally activated processes that would allow for energy-barrier crossings. All the important features of the behavior of the RFIM at zero temperature (hysteresis, avalanches, and out-of-equilibrium phase transition) are thus preserved. This approach is expected to be valid on an intermediate time scale during which the system can equilibrate locally but is not able to reach the global thermal equilibrium. In many disordered systems, the free-energy barriers are much larger than $k_BT$ and this intermediate time scale is the one probed by experiments.

In the next section, we describe the model and briefly present the local mean-field approach in the field-driven case. The algorithmic method for the magnetization-driven situation is presented in section III and the numerical results in the strong and weak-disorder regimes are discussed in section IV. Some concluding remarks are given in section V.

\section{H-driven dynamics for the RFIM at finite temperature} 

We consider the ferromagnetic (Gaussian) RFIM defined by the energy function
 
\begin{align}
{\cal U}(\{S_i\})= -J\sum_{<ij>} S_i S_j -\sum_i h_i S_i
\end{align}
where $S_i$  ($i=1, ..., N$) are Ising spins placed on the sites of a $3$d cubic lattice of size $L\times L\times L$ and $h_i$ are quenched random fields drawn from a Gaussian distribution with zero mean and standard deviation $\sigma$. The first sum runs over all distinct pairs of nearest-neighbor sites.

In the ``unconstrained" situation that was considered in all previous studies\cite{SDP2004}, the system is submitted to an external field $H$ that is changed adiabatically. At zero temperature, it then tries to follow the field evolution by minimizing the magnetic enthalpy ${\cal H}={\cal U}-HM$ where  $M=\sum_iS_i$ is the overall magnetization. The relaxation dynamics generally used in this case is the $T=0$ version of the Glauber single-spin-flip dynamics where each spin is aligned to its local field $f_i+H$, where 
\begin{align}
f_i= J\sum_{j/i}S_j+h_i
\end{align} 
and the sum is over the nearest neighbors of site $i$ (this corresponds to a local, {\it partial} minimization of ${\cal H}$\cite{note1}). A configuration $\{S_i\}$ is then (meta)stable according to the single-spin-flip dynamics when all the spins satisfy the stability condition
\begin{equation}
S_i=\mbox{sign}(f_i +H) \ .
\end{equation}
When $H$ is changed from an initially stable state, a spin may become unstable, which  may initiate an avalanche of other spin flips in the neighborhood. The avalanche ends when a new stable state is reached.  
The resulting hysteresis loop (after averaging over disorder and extrapolating to the thermodynamic limit) is continuous for $\sigma>\sigma_c$, with all avalanches being of microscopic size, and discontinuous for $\sigma<\sigma_c$, the discontinuity corresponding to a macroscopic avalanche that involves a finite fraction of the spins. The critical disorder for the RFIM on the cubic lattice has been numerically located at $\sigma_c \simeq 2.2$ \cite{SDP2004,PV2003}.

Extension of this behavior to finite temperature is straightforward when using a local mean-field approach. In this approximation, the Helmholtz free-energy of the system for a given realization of the random fields is a function of the thermally averaged local magnetizations $m_i$,
\begin{align}
&{\cal F}(\{m_i\})=-J\sum_{<ij>}m_im_j -\sum_i h_i m_i\nonumber\\
&+ \frac{1}{2\beta} \sum_i \Large\{(1+m_i)\ln(\frac{1+m_i}{2})+(1-m_i)\ln(\frac{1-m_i}{2})\Large\}
\end{align}
where $\beta=1/(k_BT)$\cite{note0}. In presence of an external field $H$, one  considers the Gibbs free-energy 
\begin{equation}
{\cal G}(\{m_i\},H)={\cal F}(\{m_i\})-H M
\end{equation}
where $M=\sum_i m_i$. The minima of ${\cal G}$ define the metastable states and satisfy the stationary conditions $\delta {\cal G}/\delta m_i=0$, i.e.
\begin{equation}
m_i=\tanh[\beta (f_i+H )]
\end{equation}
where 
\begin{equation}
f_i= J\sum_{j/i}m_j+h_i
\end{equation}
(in the following, the coupling constant $J$ is taken as the energy unit and set to unity). The method then consists in following the evolution of  the  system among these local minima as they are displaced or destroyed when the temperature $T$ or the field $H$ is changed\cite{SLG1983}.  Since one forbids any barrier crossing, the system is forced to stay in the same miminum until the stability limit is reached (which is signaled by the vanishing of some eigenvalue of the Hessian matrix $\partial^2 {\cal F}/ \partial m_i \partial m_j$). The system then jumps instantaneously to another nearby minimum. At constant temperature,  this jump defines an $H$-driven avalanche. The dynamics has thus the same irreversible and hysteretic character as the $T=0$ single-spin-flip dynamics\cite{SDKKRS1993}. The only effect of temperature is to change the shape  of the free-energy landscape. In particular, there is only one minimum at high temperature and the evolution becomes reversible. 

In practice, Eqs. (6) and (7) are solved iteratively (i.e. $m_i^{(n+1)}=\tanh[\beta (\sum_{j/i}m_j^{(n)}+h_i+H )]$ where $m_i^{(n)}$ is the value of $m_i$ after $n$ iterations), changing $H$ by small increments and using as  initial configuration  the $m_i$'s obtained with the preceding value of $H$. A crucial property of such an iterative scheme is that it only converges to minima of the free-energy (and not to maxima or saddle-points)\cite{BM1979,LBL1983}. Thanks to the convexity of the  hyperbolic tangent function in Eqs. (6), this  ``dynamics" has also the same properties as the $T=0$ dynamics: it is abelian\cite{SSD2000} (i.e. the metastable state reached after an avalanche does not depend upon the order in which the local magnetizations are updated during the avalanche) and it satisfies return-point memory\cite{SDKKRS1993}.

\begin{figure}[htb]
\begin{center}
\epsfig{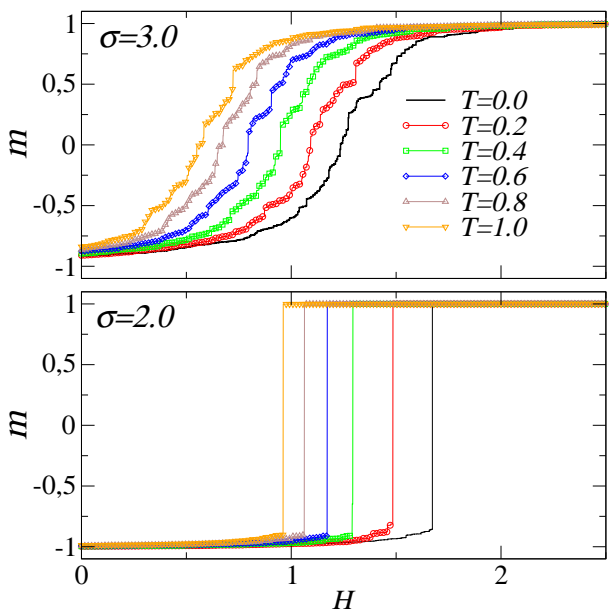}
\end{center}
\caption{Ascending branch of the $H$-driven hysteresis loop computed within local mean-field theory at different temperatures in a system of size $L= 20$. The $T=0$ curve is obtained using the single-spin-flip algorithm as in Ref.\cite{SDKKRS1993}.(Color on line)}
\end{figure}

It is clear that the Gibbs free-energy obtained from Eq. (5) identifies with the exact enthalpy ${\cal H}={\cal U}-HM$ when $T\rightarrow 0$  and $m_i\rightarrow s_i=\pm1$. The $N$ coupled equations (6) then reduce  to the stability conditions given by Eqs. (3).
The iterative dynamics at finite temperature thus reduces to the usual single-spin-flip metastable dynamics when $T\rightarrow 0$. In particular, the hysteresis loop goes to the same $T=0$ loop, as illustrated in Fig. 1 below and above the critical disorder (in this figure and in the following, the temperature unit is $J/k_B$. The mean-field critical temperature of the pure system is thus $T_c=6$).
On the other hand, since thermally activated processes are completely neglected, the $T=0$ critical point where a macroscopic avalanche appears for the first time in the thermodynamic limit is now replaced by a {\it full} line $\sigma_c(T)$ of out-of-equilibrium critical points\cite{DKRT2005}, as shown schematically in Fig. 2 (note that a precise numerical determination of the critical line would require an extensive finite-size scaling study as in Ref.\cite{DKRT2005}). Moreover, since the theory fully accounts for the disorder-induced fluctuations that play a dominant role in random-field systems, the critical behavior along the line is expected to be non-classical and identical to the one observed at $T=0$\cite{DKRT2005}. 
\begin{figure}[htb]
\begin{center}
\epsfig{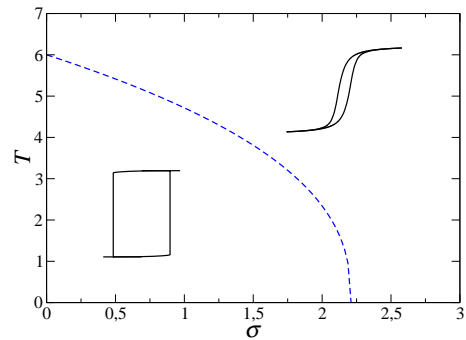}
\end{center}
\caption{Schematic out-of-equilibrium phase diagram of the $H$-driven RFIM in the local mean-field approximation. The critical line $\sigma_c(T)$ separates the  weak and strong disorder regimes characterized respectively by the presence or absence of a macroscopic jump in the hysteresis loop. (Color on line)}
\end{figure}

\section{M-driven dynamics at finite temperature: algorithm and numerical method}

We now consider the ``constrained" situation where the magnetization of the system is changed externally and the field becomes an output variable. For simplicity, we assume that the magnetization $m=M/N$ is monotonously increased  from $-1$ to $+1$ (generalization to more complicated magnetization histories is straightforward). In this case, the system tries to minimize its Helmholtz free-energy ${\cal F}$ instead of ${\cal G}$ while satisfying the global constraint $\sum_i m_i=M$. 

Since there is no intrinsic dynamics for Ising spins, there is no obvious way to define an $M$-driven dynamics for the RFIM that would be the counterpart of the $T=0$ single-spin-flip dynamics used in the $H$-driven case. In fact, it turns out that the simplest possible choice for a $T=0$ dynamics  leads to an hysteretic behavior of the system that does not allow for dissipation, which does not seem very realistic\cite{IRSV2006}. On the other hand, at finite temperature within the local mean-field approach, one is dealing with a standard problem of constrained optimization that can be solved in a natural way  by the method of Lagrange mutipliers. One first introduces the function
\begin{equation}
{\cal L}(\{m_i\},\lambda)={\cal F}(\{m_i\})-\lambda\{\sum_i m_i -M\}
\end{equation}
where $\lambda$ is a Lagrange multiplier that has the meaning of a field coupled to the extensive variable $\sum_i m_i$. Minimizing ${\cal F}(\{m_i\})$ with the constraint on the magnetization  amounts to solve simultaneously the $N+1$ coupled equations $\partial {\cal L}/ \partial m_i=0$ and $\partial{\cal L} / \partial \lambda=0$, 
 
\begin{equation}
m_i=\tanh[\beta (f_i+ \lambda)]
\end{equation}
and
\begin{equation}
\sum_i m_i(\lambda) =M \ .
\end{equation} 
In a second step, one has to define an iterative scheme that tells the system how to go from one solution to the ``nearest" one as the magnetization is slowly changed. For that purpose,  we rewrite Eqs. (9) as
\begin{equation}
1+m_i=e^{2\beta \lambda}e^{2\beta f_i}(1-m_i)
\end{equation}
and sum over $i$ so to express $\lambda$ as a function of $m$ and the local magnetizations,
\begin{equation}
\lambda=\frac{1}{2\beta} \ln\frac{1+m}{\frac{1}{N}\sum_i e^{2\beta f_i}(1-m_i)} \ .
\end{equation}
Eqs. (9), (7) and (12) define a set of nonlinear coupled equations that can be solved iteratively, as detailed below\cite{note2}. By construction,  the  configurations $\{m_i^*\}$, solutions of Eqs. (9), are extrema of the Gibbs free-energy for the special value of the field $\lambda^*(m)$ that satisfies the constraint equation, Eq.(10).
In other words, the field $\lambda$ adjusts itself in the course of the iterations so that the system eventually reaches an extremum of ${\cal G}$ with the correct magnetization. One can imagine that this is indeed the role of the feedback control used in magnetic materials\cite{B1998}. It is worth noticing that the constraint $\sum_i m_i=M$ is not satisfied until convergence is reached, which of course is the very principle of the Lagrange method. It is therefore unclear if one can attribute a physical meaning to the intermediate stages of the iteration, although a similar problem may also occur in experiments\cite{note3}.
  
In practice, to improve the convergence of the iteration procedure, we introduce a mixing parameter $0<\alpha\leq 1$ that controls how much of the previous iteration is retained. When changing the magnetization from $m$ to $m+\delta m$, we thus search for a fixed point of the map

\begin{align}
m_i^{(n+1)}&=\alpha\tanh[\beta (f_i^{(n)}+\lambda^{(n)})]+(1-\alpha)m_i^{(n)}\nonumber\\
\lambda^{(n+1)}&=\frac{\alpha}{2\beta} \ln\frac{1+m}{\frac{1}{N}\sum_i e^{2\beta f_i^{(n)}}(1-m_i^{(n)})}+(1-\alpha)\lambda^{(n)}
\end{align} 
applied to all $m_i$ in parallel. Convergence is assumed when max$(\vert m_i^{(n+1)}-m_i^{(n)}\vert,2\beta \vert \lambda^{(n+1)}-\lambda^{(n)}\vert)<\epsilon$ (with $\epsilon=10^{-4}$ in most of the calculations). 
\begin{figure}[hbt]
\epsfig{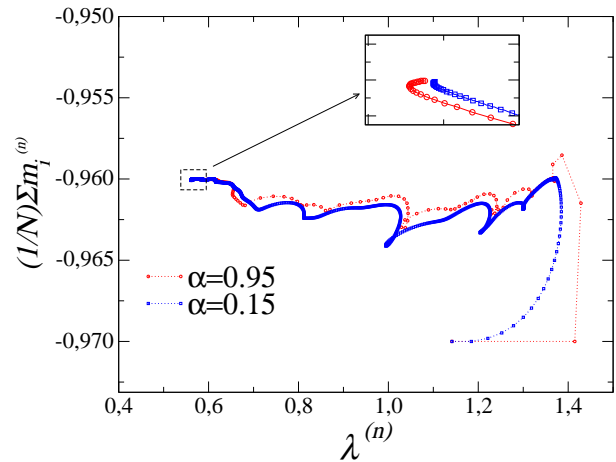}
 \caption{Typical evolution of the system in the magnetization-field plane  during the iteration procedure for two values of the mixing parameter $\alpha$ ($\sigma=1, T=2, L=20)$. $m$ is varied from $-0.97$ to $-0.96$ ($\delta m=0.01$). The inset is a magnification of the behavior during the last iterations. (Color on line)}
\end{figure}

The following general observations can be made:

	(i) In contrast with the $H$-driven situation, there is no obvious relation between the stability of the iterative scheme and the properties of the Hessian matrix\cite{LBL1983}. However, we strongly believe that convergence to a fixed point of the map (13) only occurs when the  final configuration $\{m_i^*\}$ is a local minimum of ${\cal G}$, i.e. when  $\partial^2 {\cal F}/ \partial m_i \partial m_j$ is positive definite. This was checked in small systems (up to $N=1000$) by diagonalizing the Hessian for each solution $\{m_i^*(m),\lambda^*(m)\}$, using a standard diagonalization routine\cite{PFTV1992}: in every case we found that all the  eigenvalues were positive within numerical accuracy. In the following, we shall therefore assume that the configurations visited by the dynamics are always metastable states. This, however, does not mean that these states are the same as those reached with the field driving, as will be discussed below.
\begin{figure}[hbt]
\epsfig{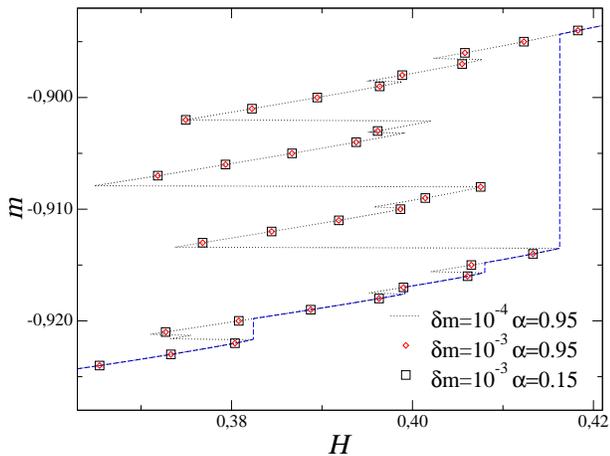}
 \caption{Portion of the $H(m)$ curve computed in a sample of size $ L=20$ for $\sigma=2$ at $T=2$, increasing $m$ by very small increments $\delta m$ and changing the value of the mixing parameter $\alpha$. The dashed (blue) curve is the corresponding $m(H)$ curve obtained when driving the system with the field. (Color on line)}
\end{figure}
	
	ii) Convergence does not occur at low temperatures (depending on $\sigma$ and $L$). The iterative map either enters limit cycles that cannot be broken by further decreasing the value of $\alpha$, or the constraint $\sum_i m_i(\lambda) =M$ starts to be violated, whatever the value of $\epsilon$. This latter problem is obviously related to the fact that some $m_i$ become equal to $\pm 1$ within numerical accuracy. The problem could perhaps be cured by working with transformed variables (see e.g. Ref.\cite{ABBM2006}) but we have not tested such a modified procedure. Of course, another possibility  is that there are no more solutions to Eqs. (9), (7), and (12) at low temperatures. We shall come back to this problem in section V.
	
	iii) The Helmholtz free-energy ${\cal F}(\{m_i\})$ may increase during the iteration procedure. This is in contrast with what happens in the $H$-driven case where it is found empirically that ${\cal G}(\{m_i\})$ always decreases during an avalanche (recall that $H$ is kept constant during an avalanche). This is perhaps an indication that the transient behavior described by the map (13) does not correspond to any reasonable physical dynamics on the free-energy landscape. But this may be a too pessimistic statement.
	
	iv) The field $\lambda$ varies in a complicated way during the iterations, as shown in Fig. 3. Accordingly, 
the evolution of the local magnetizations with $m$ is not necessarily monotonous  and the numerical results may depend on the choice of the increment $\delta m$ and/or the mixing parameter $\alpha$.  In principle, this could be a serious shortcoming of the iterative scheme implying that the corresponding dynamics is not deterministic (or ``adiabatic" in the sense of Ref.\cite{SDKKRS1993}), i.e., the final state depends on how the map (12) is applied. Note for instance that the final values of $\lambda$ shown in the inset of  Fig. 3 slightly differ for  $\alpha=0.95$ and $\alpha=0.15$. However, there is good numerical evidence that an adiabatic behavior is recovered when $\delta m$ is small enough. This is illustrated in Fig. 4 showing a portion of the curve $H(m)$ computed in a single disorder realization of size $L=20$ for two values of $\delta m$ and  $\alpha$ (in order to facilitate the comparison with the $H$-driven  magnetization curve, we now use the same notation $H$ for the field $\lambda^*(m)$ and plot the curve $H(m)$ with $m$ on the vertical axis).  One can see that the trajectory is independent of $\alpha$ for $\delta m=10^{-3}$. Moreover, the same final states are obtained with $\delta m=10^{-3}$ or $10^{-4}$. Unfortunately, one cannot compute a whole curve in a reasonable time using such small values of $\delta m$ and most of the calculations presented in the following were computed with $\delta m=10^{-2}$. With this value, the dynamics is not fully adiabatic but the variations when changing $\alpha$ are reasonably small, as illustrated in Fig. 3.

\section{Results}

The main results of our study are summarized in Fig. 5 where we compare the $m(H)$ and $H(m)$ hysteresis loops obtained  with the $H$-driven and $M$-driven dynamics, respectively, for three different values of $\sigma$ at $T=2$(well below the mean-field critical temperature of the pure system). These calculations were performed in samples of linear size $L=60$ or $L=80$\cite{note5}.
One can readily notice two remarkable features: i) the $M$-driven hysteresis loops display strong fluctuations with $m$, especially at low  $\sigma$, and ii) the behavior is quite different in the strong and weak-disorder regimes characterized respectively by the absence or presence of a macroscopic ($H$-driven) avalanche. (Although we do not know the exact  critical line $\sigma_c(T)$, the changes in the shape of the $m(H)$ curve in Fig. 5 clearly indicate that $2<\sigma_c(T=2)<3$).

\begin{figure}[hbt]
\epsfig{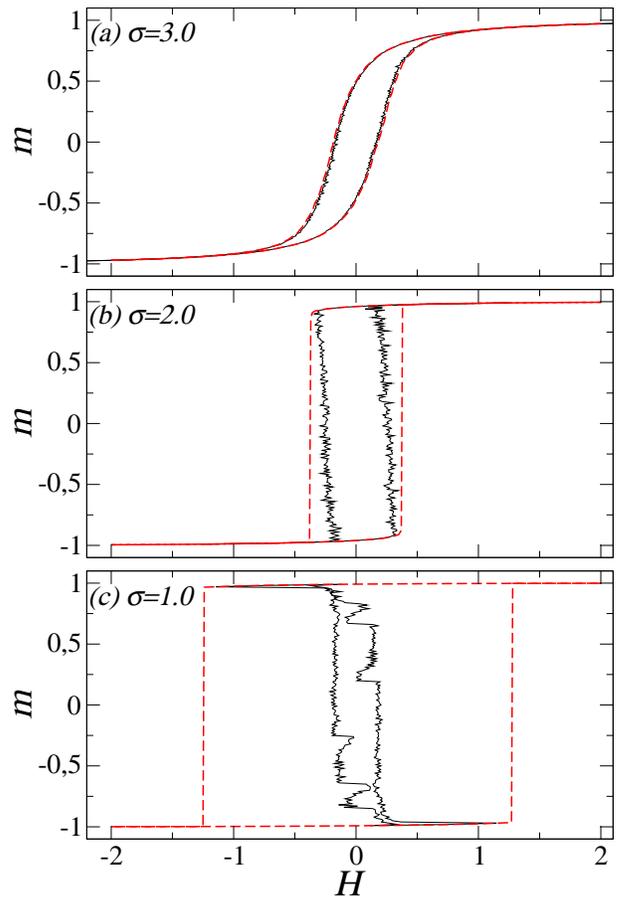}
 \caption{Comparison of the hysteresis loops obtained in a single sample with the  $H$-driven (dashed line) and  $M$-driven (solid line) algorithms for three values of $\sigma$ at $T=2$. The size of the sample is  $L=60$ (for $\sigma=3$) or $L=80$ (for $\sigma=2$ and $1$). (Color on line)}
\end{figure}

Before discussing these results in more detail, let us first comment on  the mechanism that underlies the behavior of the induced field as one slowly increases the magnetization. 
At the beginning, when $m$ is close to $-1$, $H$ has a large negative value and follows $m$ smoothly because there exists only one stable state, the equilibrium one. Hence the two curves $H(m)$ and  $m(H)$ are identical. Then, for a certain value of $m$, there is a discontinuous jump in the field towards a lower value. This corresponds to the first avalanche in the $m(H)$ curve. One can thus associate this jump to the disappearance of the initial minimum: the system has to find another metastable state with the new  magnetization. One then again follows this mimimum smoothly, until it reaches its own stability limit. Then, there is a new jump to the left and the evolution proceeds like this until $m=+1$. A similar behavior is observed along the descending branch. As can be seen in Fig. 4, when increasing $m$, the excursions of the field are always to the {\it left} of the $H$-driven magnetization curve and there may be several discontinuities in $H(m)$ corresponding to a single avalanche in $m(H)$. Indeed, because of the ferromagnetic nature of the interactions, one can prove (at $T=0$\cite{DRT2005}, but also at finite temperature in the framework of local mean-field theory) that {\it all} the metastable states are inside the $H$-driven saturation loop.

If this interpretation is correct, the magnitude of the horizontal jumps in $H(m)$ is thus related to the organization and the number of metastable states inside the $H$-driven hysteresis loop. This explains the two different behaviors observed above and below the critical disorder $\sigma_c(T)$.

\subsection{Strong-disorder regime}

\begin{figure}[hbt]
\epsfig{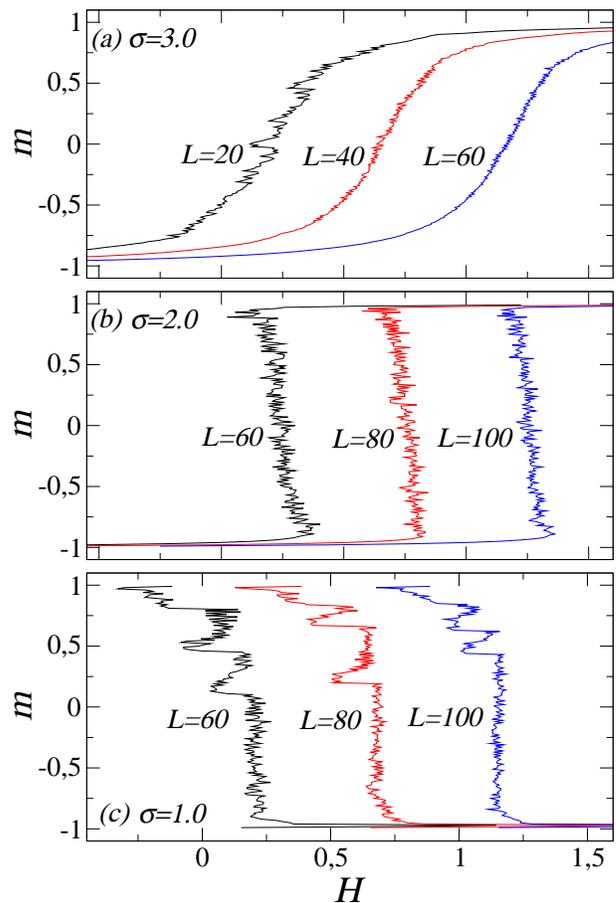}
 \caption{Evolution of the ascending branch of $H(m)$ with system size for $\sigma=3,2$ and $1$ at $T=2$. For the sake of clarity, the curves for the two larger sizes are displaced horizontally by $0.5$ and $1$.(Color on line)}
\end{figure} 

We first consider the strong-disorder regime $\sigma>\sigma_c(T)$. 
As can be seen in Fig. 5(a), the fluctuations in $H(m)$ are small, which may be traced back to the fact that  there are many metastable states in the vicinity of the $H$-driven magnetization curve. More precisely, one expects the number of metastable states inside the hysteresis loop to scale exponentially with $N$ when $\sigma>\sigma_c(T)$. This was proven analytically  in Ref.\cite{DRT2005} for the $1$-d RFIM at $T=0$ and is probably always true in the strong-disorder regime where the saturation loop is continuous. In particular, Fig. 4 in Ref.\cite{DRT2005} shows that the complexity, i.e. the logarithm of the number of metastable states divided by $N$,  increases very rapidly in the vicinity of the boundary (whereas it is zero outside the loop, indicating that there are no metastable states in this region when $N \rightarrow \infty$).
Since, moreover, the $H$-driven avalanches remain of microscopic size when $\sigma>\sigma_c(T)$, we expect the small wiggles in the $H(m)$ curve to become infinitesimally small in the thermodynamic limit. This is indeed what is observed in Fig. 6(a) where the curves obtained for different system sizes are compared. Therefore, we predict that the two curves $H(m)$ and  $m(H)$ should  become identical in the thermodynamic limit, as suggested by Fig. 5(a).
\begin{figure}[hbt]
\epsfig{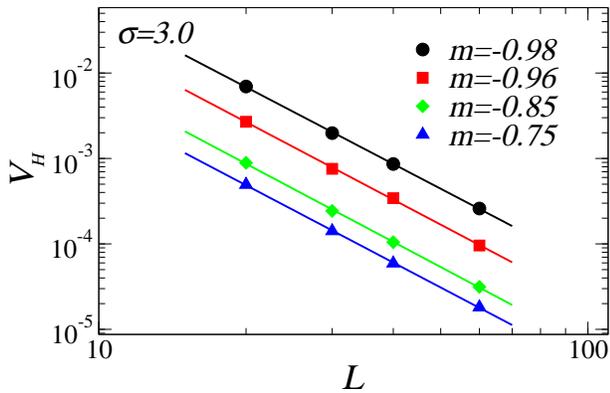}
 \caption{Variance $V_H(m)$ of $H(m)$  for selected values of $m$  as a function of system size  for $\sigma=3$ and $T=2$. The number of samples  is $10000$ for $L=20$ and $1000$ for $L=30, 40$ and $60$.  For the sake of clarity, the variances for $m=-0.96, -0.85$ and $-0.75$ are divided by $2, 4$ and $6$, respectively. The lines are fits to the form $V_H(m)\sim L^{-\rho}$, yielding $\rho \approx 3 $ in all cases. (Color on line)}
\end{figure} 

Note that these heuristic and numerical arguments imply that $H(m)$ is a self-averaging quantity for $\sigma>\sigma_c(T)$. $H(m)$ has a different value for each realization $\{h_i\}$ of the random fields and the issue of self-averaging concerns the behavior of the width of the probability distribution $P_L(H(m))$ as $L$ increases. If $H(m)$ is self-averaging, ``most" realizations (in a sense that can be made precise) lead to the same value of $H(m)$ in the thermodynamic limit.  If not, the measurement of the field in a single sample, no matter how large, does not give a meaningful result and must be repeated on many samples. The situation is clearly different from the usual $H$-driven case where the magnetization is the output variable: $H(m)$ is indeed an {\it intensive} quantity and imposing the magnetization $M$ of a large sample does not impose the magnetization in large subsamples. Therefore, the standard Brout argument\cite{B1959} cannot be applied and one cannot say anything {\it a priori} about the variations of the variance $V_H(m)=\langle H(m)^2\rangle-\langle H(m)\rangle ^2$ with $L$ (hereafter, $\langle ...\rangle$ denotes the quenched average over disorder). However, if the two curves $m(H)$ and $H(m)$ are indeed the same in the thermodynamic limit, $H(m)$ is a self-averaging quantity, like $m(H)$. This is confirmed by Fig. 7 that shows that the variance $V_H(m)$ does vary like $L^{-3}$ (note that this calculation has only been performed in the initial part of the curve, in the range $-1<m<-0.75$, so to reduce the computational effort).

\subsection{Weak-disorder regime}

\begin{figure}[hbt]
\epsfig{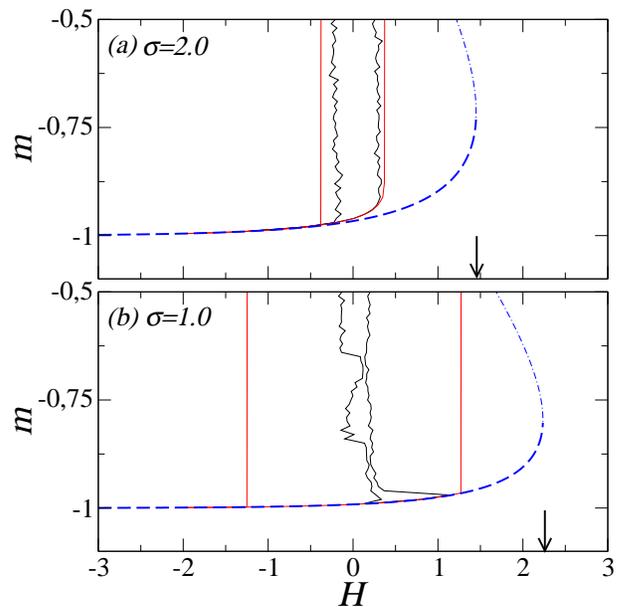}
\caption{Magnification of the lower part of Figs. 5(b) and 5(c). The dashed and dotted-dashed lines represent the metastable and unstable branches of the global mean-field curve obtained from Eq. (14). The arrows indicate the field $H_{sp}(\sigma)$ at the stability limit. (Color on line)}
\end{figure}
We now consider the case $\sigma<\sigma_c(T)$ illustrated in Figs. 5(b) and 5(c). There are two striking differences with Fig. 5(a): i) the curve $H(m)$ shows a pronounced re-entrant behavior, and ii) the field fluctuates much strongly with $m$. In line with the preceding discussion, we attribute the re-entrant behavior to the peculiar distribution of the metastable states in the field-magnetization plane in the weak-disorder regime. This issue was discussed in Ref.\cite{DRT2005} where it was suggested (but not proven) that the macroscopic jump in the $m(H)$ curve is related to the absence of metastable states in a certain region of the field-magnetization plane (see Fig. 10 in that reference). If this scenario is correct, the field, in the $M$-driven situation, has to jump to the other side of this region so to find metastable states with the correct magnetization. We thus suspect that the $H(m)$ curve is close to (or may be, in the thermodynamic limit, identical to) the boundary of this region. 
We want to stress that the re-entrant behavior discussed here has nothing to do with the $S$-shape predicted by the {\it global} mean-field theory where all the $m_i$'s are taken equal to the average magnetization $\langle m \rangle$. In Figs. 8(a) and 8(b), the lower part of the curves $H(m)$ are compared to the mean-field prediction obtained by solving the implicit equation\cite{SP1977}
\begin{align}
\langle m \rangle=\int_{-\infty}^{+\infty} \tanh [\beta(z \langle m \rangle+h+H)] P(h) dh
\end{align} 
where $z=6$ is the lattice coordination number and $P(h)$ is the Gaussian probability distribution. As can be seen in Fig. 8(a) for $\sigma=2$, when increasing $m$, the field $H(m)$ deviates from the mean-field curve  much before the spinodal $H_{sp}(\sigma)$ is reached and it starts to strongly fluctuate with $m$ just at the onset of the macroscopic avalanche in $m(H)$ (note that the intermediate part of the mean-field curve is {\it unstable} whereas all the states visited by the $M$-driven dynamics are {\it metastable}). Clearly, an average description cannot properly describe the metastable behavior associated to the existence of {\it local} inhomogeneities in the system (it is only when  $m$ is close to $\pm 1$ and the system is quasi-homogeneous that the local and global descriptions become similar).

The same is true for $\sigma=1$ in Figs. 5(c) or 8(b), but the re-entrant behavior is much more pronounced and the field almost drops to zero. Indeed, one expects the
metastable states to concentrate in the vicinity of $H=0$ as the disorder  decreases (see for instance Fig. 1 in Ref.\cite{DRT2005}). When $\sigma=0$, the only metastable states that remain in non-zero field are the two {\it extremal} ones that correspond to the metastable branches of the mean-field curve for the pure system, solution of $m=\tanh[\beta(cm+H)]$. Therefore, in the limit $\sigma \rightarrow 0$, as $m$ is increased from $m=-1$,  the field $H(m)$ follows the lower  metastable branch up to $H_{sp}(\sigma=0)$, then drops to zero and stays to this value until the upper metastable branch is recovered and followed up to $m=+1$. The intermediate metastable states in zero field then correspond to bubbles of the minority phase that grow progressively as $m$ is increased (controlling $m$ is like controlling the spinodal decomposition of the system).  When $\sigma$
is {\it finite} but small, one expects a similar scenario: initially, the system is able to follow the imposed magnetization while remaining quasi-homogeneous; then, for a certain value of $m$ (and a corresponding value of $H$ that is {\it smaller} than $H_{sp}(\sigma)$ as can be seen in Figs. 8(a) and 8(b)), this situation becomes unstable and a nonequilibrium multi-domain state is formed while the field decreases suddenly; this  multi-domain state  then evolves progressively until the system becomes homogeneous again.
This scenario may be compared with the one presented in Ref.\cite{B1998} (see p. 321 in that reference) to explain the re-entrant behavior observed in picture-frame single crystals when the rate of change of the magnetic flux is imposed by using a feedback control\cite{S1950}. In this case, the motion of a  single domain wall is the dominant process. However, one can notice the similarity of the curve plotted in Fig. 5(c) with the experimental curve\cite{S1950}. A similar behavior is also observed in strain-stress curves measured in shape memory alloys under displacement control loading condition (see e.g. Refs.\cite{LS2002,BMPRV2006}). In this case, the physical origin of the sudden drop in the stress after it has reached a critical value is still debated (in the recent simulations of  Ref.\cite{ALS2006}, it is  associated to the formation of a multi-domain state, like in the present work). Our results suggest that this is probably a generic behavior of the hysteresis loops obtained in weakly disordered systems when controlling the extensive variable.

\begin{figure}[hbt]
\epsfig{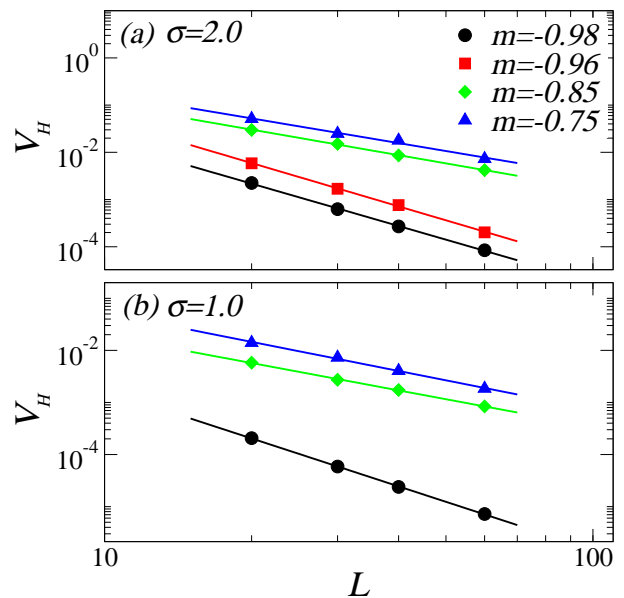}
\caption{Same as Fig. 7  for $\sigma=2$ and $\sigma=1$.  For the sake of clarity, the variances for $\sigma=2$ and $m=-0.96, -0.85, -0.75$ are mutiplied by $2, 4$ and $8$, respectively.  The variance for $\sigma=1$ and $m=-0.96$ is not shown, as it does not decrease with $L$ (see text) and the one for $m=-0.75$ is multiplied by $3$. The fit of the variance to the form $V_H(m)\sim L^{-\rho}$ yields $\rho$ varying from $3$ to  $\simeq 1.75 $ as $m$ increases from $-0.98$ to $-0.75$. (Color on line)}
\end{figure}
We now move to the discussion of the discontinuous fluctuations of the field $H$ with the magnetization.
As shown in Figs. 6(b) and 6(c), the magnitude of these fluctuations decrease with system size but they are still very significant for $L=100$, the largest size investigated in the present work (we note however that the strongest discontinuities observed in Fig. 6(c) are displaced to larger and larger values of $m$ as $L$ increases, which suggests that they  disappear as $L \rightarrow \infty$).  One may wonder whether one should observe such a sporadic, discontinuous behavior of the induced field in a macroscopic sample. This is again related to the issue of self-averaging.
As shown in Figs. 9(a) and 9(b) (for selected values of $m$ in the lower part of the $H(m)$ curve), we find that the variance $V_H(m)$ in the weak-disorder regime still decreases with $L$, but the finite-size scaling exponent becomes significantly smaller than $3$ ($\rho \simeq 1.75$) as soon as $m$ is in the range of the macroscopic avalanche. If this behavior does not change for larger $L$, this seems to indicate that $H(m)$ is only {\it weakly} self-averaging\cite{BH1988}, a rather remarkable behavior for a non-critical system. On the other hand, one may note that the corresponding histograms shown in Fig. 10 are still very well fitted by a Gaussian distribution.
 \begin{figure}[hbt]
\epsfig{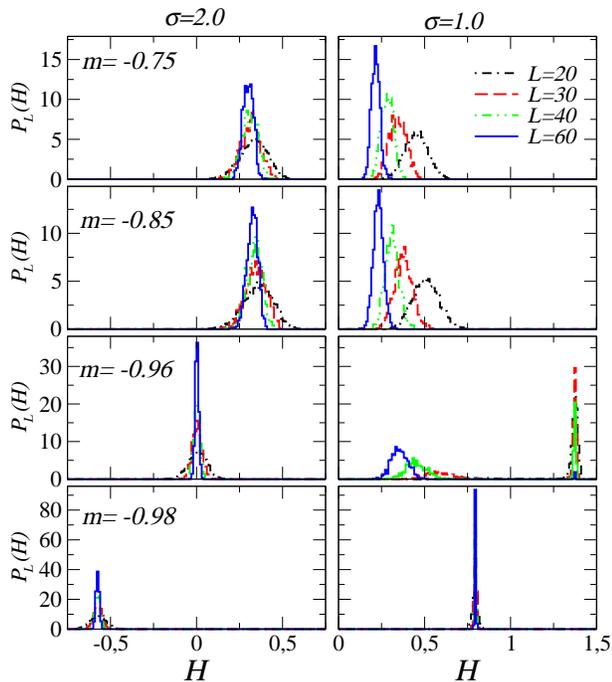}
\caption{Histograms of $H(m)$ for different values of $m$ in the range $-1<m<-0.75$ for $\sigma=1, T=2$ and different system sizes. The statistics is  performed over $10000, 1000, 1000$ and $600$ disorder realizations for $L=20, 30, 40$ and $60$, respectively.(Color on line)}
\end{figure}

What happens just at the onset of the avalanche for $\sigma=1$ deserves a separate discussion. In this case, the sample to sample fluctuations of the field are so large that the variance does not decrease at all with the system size.  Indeed,  as can be seen in the right pannel of Fig. 10 for $m=-0.96$, $H$ fluctuates between two quite distinct values, $H_{max}\approx 1.4$ and  $H_{min}\approx 0.4$ (in Fig. 10, the two peaks are simultaneously present for $L=40$ only, but a similar behavior occurs for other values of $L$ and slightly different values of $m$). This double-peak structure, which does not exist for  $\sigma=2$, could indicate the occurence of a true discontinuity in $H(m)$ in the thermodynamic limit, associated with the sudden 
appearance of the inhomogeneous multi-domain state. Rather, we believe that the presence of the two peaks is a finite-size artifact: the number of metastable states between $H_{min}$ and $H_{max}$ is probably so small that much larger statistics would be needed to observe intermediate values of the field. It is more likely that the drop in the field at the onset of the multi-domain state is very steep but that the curve $H(m)$ is nevertheless continuous. 

A similar behavior is observed for the Helmholtz free-energy ${\cal F}$ as a function of the magnetization: it is strongly self-averaging at large disorder (with a variance decaying like $1/L^3$) and only weakly self-averaging when there is a macroscopic $H$-driven avalanche. In this latter case, however, the exponent $\rho$ seems to be larger ($\rho \simeq 2.45$) than the one  associated to the fluctuations of the field. 

\section{Concluding remarks}

To summarize, we have presented a detailed study of the nonequilibrium response of the random field Ising model when one cycles adiabatically the magnetization ($M$-driven protocol) instead of the magnetic field ($H$-driven protocol).  The study has been performed in the framework of the local mean-field approach at finite temperature, neglecting all thermally activated processes. We have shown that two regimes can be observed, depending on the disorder strength. In the strong disorder regime, where all $H$-driven avalanches are of microscopic size, the two protocols yield the same hysteresis loop in the thermodynamic limit. On the other hand, in the weak-disorder regime, where the $H$-driven magnetization curve is discontinuous, the $M$-driven hysteresis loop is re-entrant and the output field displays strong fluctuations with the magnetization. There are also considerable sample-to-sample fluctuations that decrease only weakly with system size.

From the results that have been presented above, a question naturally arises: what is the behavior of the system as the temperature is decreased to zero ? As noted in the introduction, temperature was introduced in the model so to work with continuous variables instead of Ising spins and to use the Lagrange constrained optimization method. The free-energy landscape changes with $T$ (which may be very important for the actual physical behavior, as shown by the study of fluid adsorption in porous solids\cite{DKRT2005}) but the dynamics is essentially a zero-temperature dynamics since all metastable states have an infinite life-time and free-energy barriers cannot be crossed. All the preceding calculations were thus performed rather arbitrarily at the reduced temperature $T=2$ (below the mean-field critical temperature of the pure system), as we were mainly interested in the qualitative behavior of the system as a function of disorder strength. The problem is that the limit $T\rightarrow 0$ seems to be non-trivial. On the one hand, one would expect a gradual evolution with $T$, as is the case with the $H$-driven protocol (see Fig. 1). As an illustration, we show in Fig. 11 the evolution of the ascending branch of $H(m)$ as $T$ is decreased from $2$ to $1$. Not surprisingly, the fluctuations with $m$ increase as $T$ decreases but one could imagine that it is possible to decrease the temperature down to zero. This optimistic view is reinforced by the hope that the envelope of the metastable states in the $H-M$ plane (which we think plays an important role in understanding the behavior of the $M$-driven system) has  a well-defined limit at $T=0$, as discussed in Ref.\cite{DRT2005}. 
\begin{figure}[hbt]
\includegraphics[width=8cm,clip]{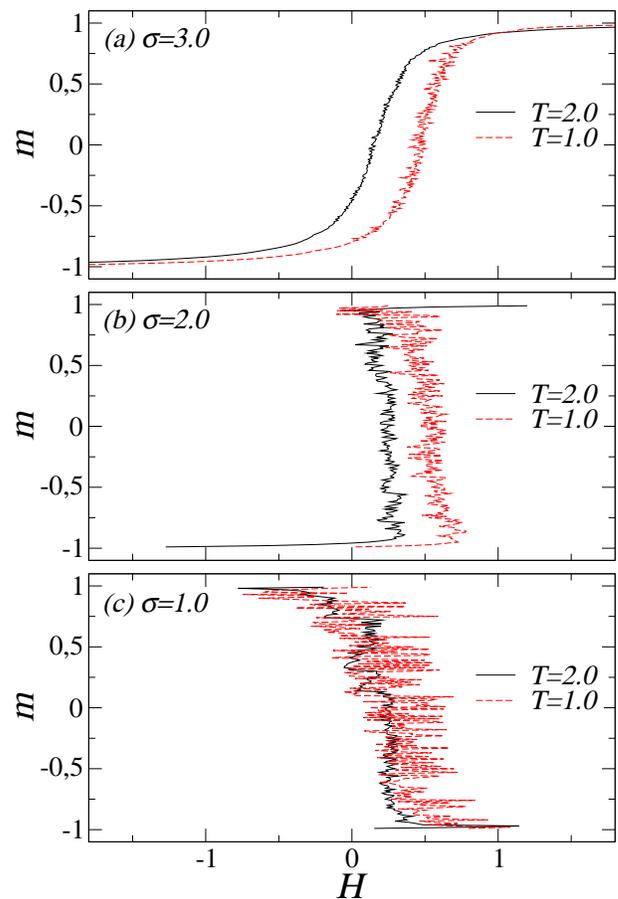}
\caption{ Effect of temperature of the ascending branch of $H(m)$. The system size is $L=40$.(Color on line)}
\end{figure}

Unfortunately, the iteration procedure does not converge any more at lower $T$ for reasons that are not completely clear. As noted earlier, this may be a problem of numerical accuracy in the solution of the iterative map. This may also be related to the fact that the magnetization $M$ becomes a discrete variable when $T\rightarrow 0$ and can only change by an increment of $\pm 2$. It may be then very difficult for the algorithm to find a path from one metastable state to another due to the existence of these gaps in the magnetization. Clearly, it would be much more convenient to define a proper dynamics directly at $T=0$. However, as discussed elsewhere\cite{IRSV2006}, this does not seem to be an easy task either.

Finally, we would like to remark that some of the features found in the present calculations, in particular the re-entrant behavior and the intermittent fluctuations in the induced field as a function of magnetization, are observed in a variety of systems characterized by an athermal dynamics. For instance, many amorphous materials, including polymers and metallic alloys, exhibit a mechanical response to deformation (e.g. the stress induced by a quasistatic shear strain) that closely resemble the field-magnetization curves computed in this work. Similar behavior is also observed in granular materials and foams. The relation of this behavior to changes in the underlying energy landscape 
has been discussed in the literature (see e.g. Ref\cite{ML1999}). However, it seems to be the first time that a direct comparison between the different driving mechanisms is performed.

\acknowledgments

The authors are grateful to Ll. Ma\~nosa, A. Planes, F. Detcheverry, E. Kierlik, and G. Tarjus for stimulating discussions. This work has received financial support from CICyT (Spain), project
MAT2004-1291 and CIRIT (Catalonia), project 2005SGR00969.  Xavier Illa
acknowledges financial support from the Spanish Ministry of Education,
Culture and Sports. M.L.Rosinberg thanks Generalitat 
de Catalunya for financial support (2005PIV1-17) and the 
hospitality of the ECM department of the University of Barcelona.  
The Laboratoire de Physique Th\'eorique de la Mati\`ere Condens\'ee is the UMR 7600 of the CNRS.


\begin{thebibliography}{10}

\bibitem{DZ2004} For a recent review on the Barkhausen effect, see G. Durin and S. Zapperi in {\it The Science of Hysteresis}, edited by G. Bertotti and I. Mayergoyz, Elsevier (2004).
\bibitem{VOMRPP1994} E. Vives, J. Ortin, L. Ma\~nosa, I. R\`afols, R. P\'erez-Magran\'e, and A. Planes, Phys. Rev. Lett. {\bf 72}, 1694 (1994).
\bibitem{LFH1993} M.P. Lilly, P.T. Finley, and R.B. Hallock, Phys. Rev. Lett. {\bf 71}, 4186 (1993).
\bibitem{B1998} See e.g. G. Bertotti, {\it Hysteresis in Magnetism}, Academic Press, San Diego (1998) and references therein.
\bibitem{WC1990} This is only true when the volume of the bulk gas outside the porous solid (in the so-called dead volume) is negligible. See for instance A. P. Y. Wong and M. H. W. Chan, Phys. Rev. Lett., {\bf 65}, 2568 (1990).  
\bibitem{OW1998} K. Otsuka and C. M. Wayman in {\it Shape Memory Materials}, edited by K. Otsuka and C. M. Wayman, Cambridge University Press, Cambridge (1998).
\bibitem{LS2002} Z. Q. Li and Q. P. Sun, Int. J. Plasticity {\bf 18}, 1481 (2002).
\bibitem{BMPRV2006} E. Bonnot, Ll. Ma\~nosa, A. Planes, R. Romero, and E. Vives (in preparation).
\bibitem{SDP2004} J. P. Sethna, K. A. Dahmen, and O. Perkovi\'c  in {\it The Science of Hysteresis}, edited by G. Bertotti and I. Mayergoyz, Elsevier (2004).
\bibitem{IRSV2006} X. Illa, M. L. Rosinberg, P. Shukla, and E. Vives (in preparation).
\bibitem{SLG1983}  C. M. Soukoulis, K. Levin, and G. S. Grest, Phys. Rev. B, {\bf 28}, 1495 (1983); G. S. Grest, C. M. Soukoulis, and K. Levin, Phys. Rev. B, {\bf 33}, 7659 (1986). 
\bibitem{note1} One could also consider relaxation dynamics that allow for multiple spin-flips, yielding a better minimization of ${\cal H}$. See for instance  E. Vives, M.L. Rosinberg, and G. Tarjus, Phys. Rev. B {\bf 71}, 134424 (2005).
\bibitem{PV2003} F. J. Perez-Reche and E. Vives, Phys. Rev. B {\bf 67}, 134421 (2003).
\bibitem{note0} This expression may be considered as the first-order term in a systematic high-temperature expansion of the free-energy at fixed local magnetizations $m_i$. See  A. Georges and J. Yedidia, J. Phys. A, {\bf 24}, 262 (1991). 
\bibitem{SDKKRS1993} J. P. Sethna, K. Dahmen, S. Kartha, J. A. Krumhansl,
B. W. Roberts, and J. D. Shore, Phys. Rev. Lett. {\bf 70}, 3347
(1993).
\bibitem{BM1979} A. J. Bray and M. A. Moore, J. Phys. C: Solid State Phys., {\bf 12}, L441 (1979).
\bibitem{LBL1983} D. D. Ling, D. R. Bowman, and K. Levin, Phys. Rev. B, {\bf 28}, 262 (1983).
\bibitem{SSD2000} S. Sabhapandit, P. Shukla, and D. Dhar, J. Stat. Phys.  {\bf 98} 103 (2000).
\bibitem{DKRT2005} For an illustration of this statement, see F. Detcheverry, E. Kierlik, M. L. Rosinberg, and G. Tarjus, Phys. Rev. E 72, 051506 (2005). In this work, one considers a model that is closely related to the RFIM and that describes the hysteretic behavior of a fluid adsorbed in a disordered porous matrix.
\bibitem{note2}  Instead of Eq. (12), one can also use the alternative expression obtained via the transformation $\lambda \rightarrow-\lambda$, $m \rightarrow-m$, $m_i\rightarrow-m_i$, $f_i\rightarrow-f_i$. The  calculation must then be started with all the $m_i$ set to $1$.
\bibitem{note3} This is especially the case when one uses a feedback control\cite{B1998}. There probably exists a transient behavior during 
which the magnetization (or the rate of variation of $m$) is not equal to the imposed value. On the other hand, in the case of strain-driven experiments in martensites, it is difficult to imagine that the length of the sample can also fluctuate.
\bibitem{PFTV1992} W. H. Press, B. P. Flannert, S. A. Teukolsky, and W. Vetterling, {\it Numerical Recipes} Cambridge University Press, Cambridge (1992).
\bibitem{ABBM2006} T. Aspelmeir, R. A. Blythe, A. J. Bray, and M. A. Moore, preprint cond-mat/0602639.
\bibitem{note5} Convergence to an accuracy of $10^{-4}$ typically requires between $10$ and $10^{3}$ iterations in a system of size $L=60$ and  several CPU hours on a $2.4$ GHz workstation are needed to compute a whole curve $H(m)$ with an increment $\delta m=0.01$. Therefore, one cannot study very large systems, in contrast with what is done at $T=0$.
\bibitem{DRT2005} F. Detcheverry, M.L. Rosinberg, and G. Tarjus, Eur. Phys. J. B {\bf 44}, 327 (2005).
\bibitem{B1959} R. Brout, Phys. Rev. {\bf 115}, 824 (1959).
\bibitem{SP1977} T. Schneider and E. Pytte, Phys. Rev. B, {\bf 15}, 1519 (1977); A. Aharoni, Phys. Rev. B, {\bf 18}, 3318 (1978).
\bibitem{S1950} K. H. Stewart, Proc. Phys. Soc. A {\bf 63},761 (1950); G. Hellmiss and L. Storm, IEEE Trans. Magn. {\bf MAG-10}, 36 (1974). 
\bibitem{ALS2006} R. Ahluwalia, T. Lookman, and A. Saxena, Acta Mater. {\bf 54}, 2109 (2006).
\bibitem{BH1988} K. Binder and D. W. Heermann, {\it Monte Carlo Simulations in Statistical Physics}, Springer-Verlag, Berlin (1988).
\bibitem{ML1999} D. L. Malandro and D. J. Lacks, J. Chem. Phys. {\bf 110}, 4593 (1999).





\end{thebibliography}
\end{document}